\documentclass[11pt]{article}
\usepackage{amsmath,amssymb,amsthm}
\usepackage{fancyhdr}
\usepackage{graphicx}
\usepackage{caption}
\usepackage{subcaption}
\newtheoremstyle{quest}{\topsep}{\topsep}{}{}{\bfseries}{}{ }{\thmname{#1}\thmnote{ #3}.}
\theoremstyle{quest}

\usepackage{url}
\usepackage[export]{adjustbox}

\title{Experimental Design for \\ Human-in-the-Loop Driving Simulations}
\date{}
\author{Katherine Driggs-Campbell, Guillaume Bellegarda, Victor Shia, \\ S. Shankar Sastry, and Ruzena Bajcsy \\
Department of Electrical Engineering and Computer Sciences \\ \texttt{krdc@eecs.berkeley.edu}}

\begin{document}

\maketitle

\setlength{\parindent}{0pt} 
\setlength{\parskip}{2ex}

\begin{abstract}
This report describes a new experimental setup for human-in-the-loop simulations.  A force feedback simulator with four axis motion has been setup for real-time driving experiments.  The simulator will move to simulate the forces a driver feels while driving, which allows for a realistic experience for the driver.  This setup allows for flexibility and control for the researcher in a realistic simulation environment.  Experiments concerning driver distraction can also be carried out safely in this test bed, in addition to multi-agent experiments. All necessary code to run the simulator, the additional sensors, and the basic processing is available for use.
\end{abstract}

\section{Introduction}
Recently, human-in-the-loop control has been a focus of many research projects.  These projects generally consider how automation and humans can work together or in a joint environment \cite{DoD}.  However, testing these algorithms and setting up experiments is often difficult due to safety concerns and lack of realistic simulations.  In this report, the details of a new driving simulator at the University of California, Berkeley, are presented to address this issue, focusing on driving experiments.

While realistic simulators as described here are fairly standard in industry \cite{Greenberg1994,ToyotaSim}, they relatively uncommon in academia, and are generally stationary simulators.  Notable work in this area and driving simulators at various universities can be found at these sites \cite{OSUsim,ClemsonSim,Trivedi,UCSDsim,StanfordSim}.  While many of these systems provide excellent visual displays and can be used in a variety of experiments, including user interface design and psychology studies, few have the motion feedback that is provided on the platform to be in active safety system design and control as presented here.  All code referred to in this paper that is needed to run the simulator, the additional sensors, and the basic processing is available for use. \footnote{Code is available at \url{http://www.purl.org/simulator_code/}}

\pagebreak

When considering safety control applications in vehicles, one of the most difficult aspects is the human.  Humans have a tendency to be not predictable, as their actions are often dependent on their mental state.  These mental states, which could be labeled as happy, angry, sleepy, etc., are difficult to identify, especially from an engineer's perspective, and differ greatly between different people.  The project that motivated this report was considering driver distraction \cite{Vasudevan2012}.  Experimenting with distracted drivers is both difficult and dangerous due to the fact researchers cannot actively tell drivers to text while driving in a real vehicle.  To ensure safety, a computer simulation environment was implemented to gather data on driver behavior \cite{Shia2013}.  However, when obtaining feedback from the subjects, it was found that a major complaint was the lack of realism in the simulations.  To address this, a Force Dynamics Simulator \cite{ForceDynamics} was setup for real-time use with PreScan, a standard industry simulation software \cite{PreScan}, to create a realistic and safe test bed for driver experiments.

This technical report presents the work that has been done to setup this new test bed with realistic simulations that mimic real driving. The experimental setup gives an overview of each component and how each contributes to the testbed.  The work that has gone into designing sensors to monitor the driver will be discussed as well as the data collection and basic processing that is available for use in developing experiments and algorithms.  Finally, the discussion and future work conclude this report.

\section{Experimental Setup}
\begin{figure}[!h]
\centering
\includegraphics[scale=0.4]{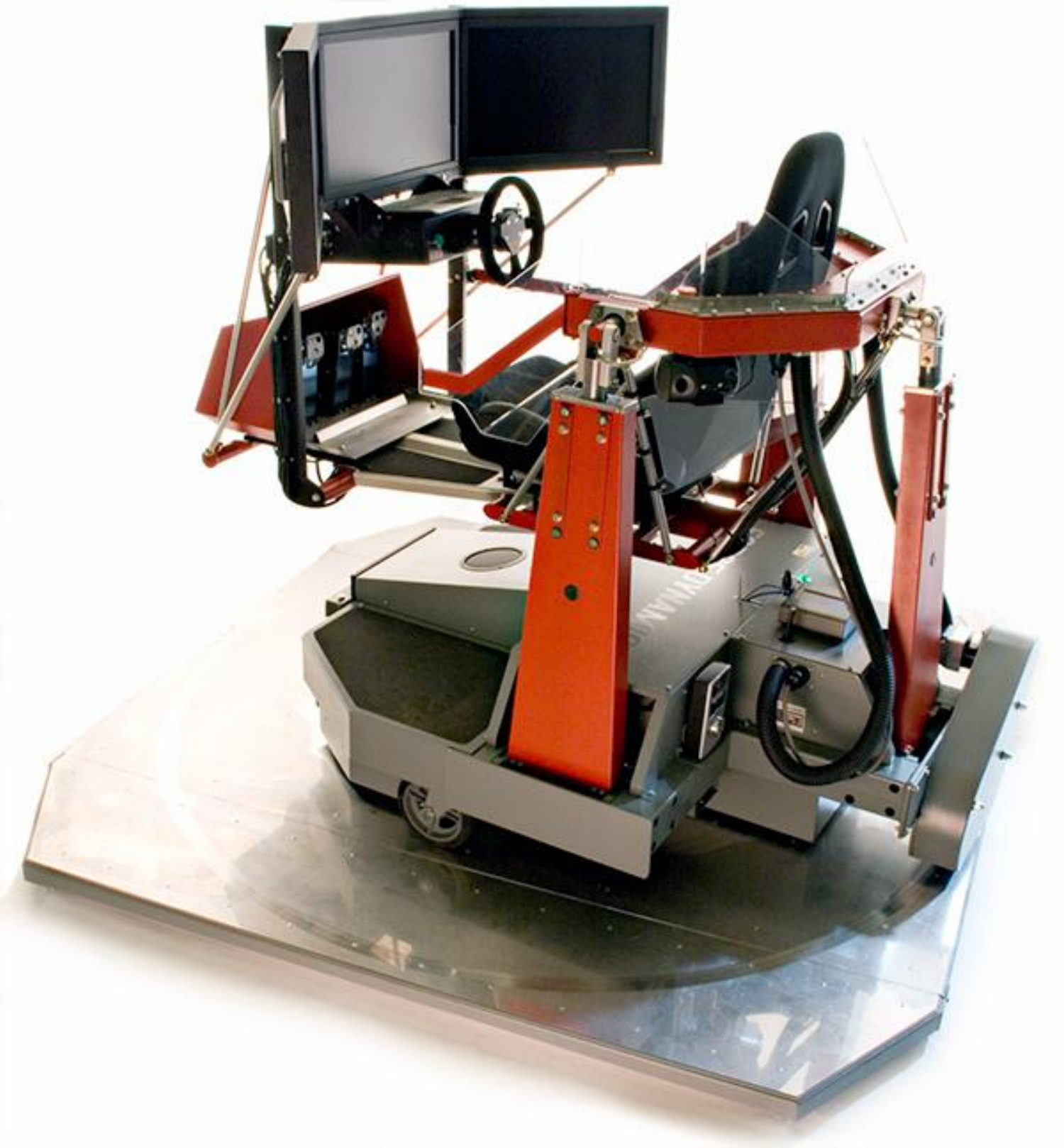}
\caption{\small Force Dynamics 401CR Platform \cite{ForceDynamics}.}
\label{fig:simulator}
\end{figure}

This section presents the simulator setup for use in human-in-the-loop experiments and driver data collection.  The simulator is a Force Dynamics 401CR platform (Fig. \ref{fig:simulator}).  The platform runs using two computers: one computer to run the software, referred to as the gaming computer, and one computer to control the movement of the simulator, referred to as the control computer.  This system has four axes: pitch, roll, yaw, and heave, which are used to simulate the forces a driver would experience.  It is equipped with a sound system and video display, which provides full frontal view with three angled screens.  For more information about the technical specifications of the simulation platform, please visit \cite{ForceDynamics}.

\subsection{Controlling the Simulator}

\begin{figure}[!h]
\centering
\includegraphics[width=\textwidth]{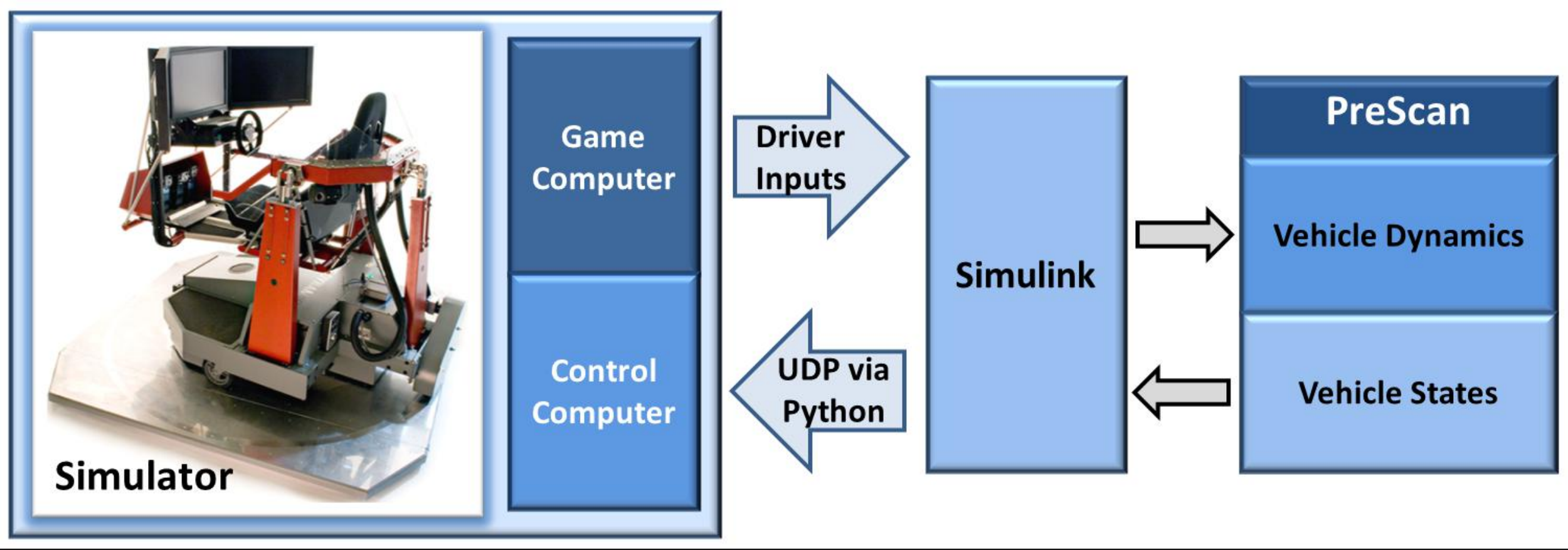}
\caption{\small Block Diagram showing the communication protocol to control the simulator.}
\label{fig:simControl}
\end{figure}

To run driving simulations, PreScan is used to design and run experiments.  While this software is often used for replaying collected data or for running experiments where real-time performance is unnecessary, a communication protocol between the software and the simulator was put in place to achieve real-time simulations with the simulator, currently running at a rate of 200 Hz.  The input to the simulator platform is obtained though the joystick function block from Matlab's Simulink, which is then used as the input to the dynamics of the vehicle through PreScan.  The position of the vehicle, in terms of pitch, yaw, and roll, is obtained from PreScan and sent to a Python script via User Datagram Protocol (UDP).  This is packaged as a specific structure that is sent to the simulator control computer, again over UDP, to adjust the position of the simulator appropriately.  This setup is illustrated in Figure \ref{fig:simControl}.  Each of these signals have gains associated them, which have been hand-tuned to re-create the feeling of driving.  While majority of the gains were selected by trial and error with multiple drivers' feedback, the braking and throttle input were adjusted such that the vehicle would decelerate from sixty to zero miles per hour in four seconds (following the idea of the four-second-rule when driving at highway speeds \cite{DMV}) and would accelerate from zero to sixty miles per hour in twelve seconds.

\subsection{Sensors to Monitor the Driver}
To gather complete data concerning driver behavior, additional components have been developed for the simulator.  Video or MS Kinect data can easily be collected by placing a camera or Kinect on the monitor of the simulator.  This can be used for pose detection or other video processing algorithms, which would give insight to driver state as shown in \cite{DriggsCampbell2013,Shia2013}.  Also, eye-tracking glasses can be worn by the driver so his exact gaze can be monitored \cite{SMI}.

\begin{figure}[!b]
\centering
\begin{subfigure}[c]{0.4\textwidth}
\includegraphics[width=\textwidth]{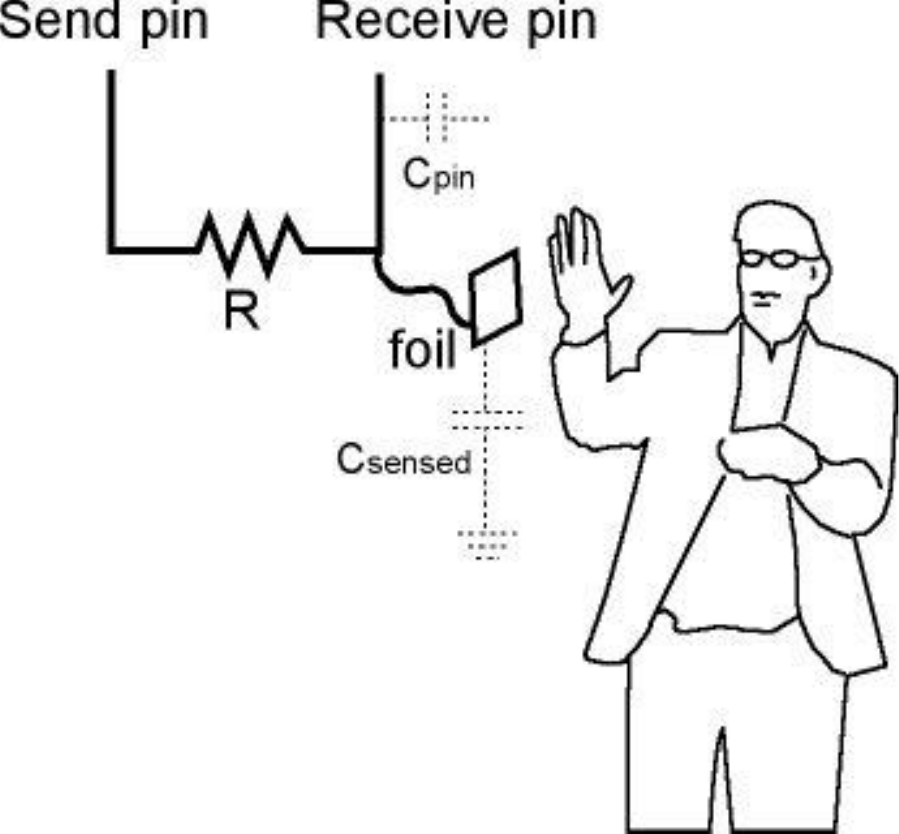}
\caption{\small Schematic for the capacitive touch sensor \cite{Badger}.}
\label{fig:sensor}
\end{subfigure}
\begin{subfigure}[c]{0.55\textwidth}
\includegraphics[width=\textwidth]{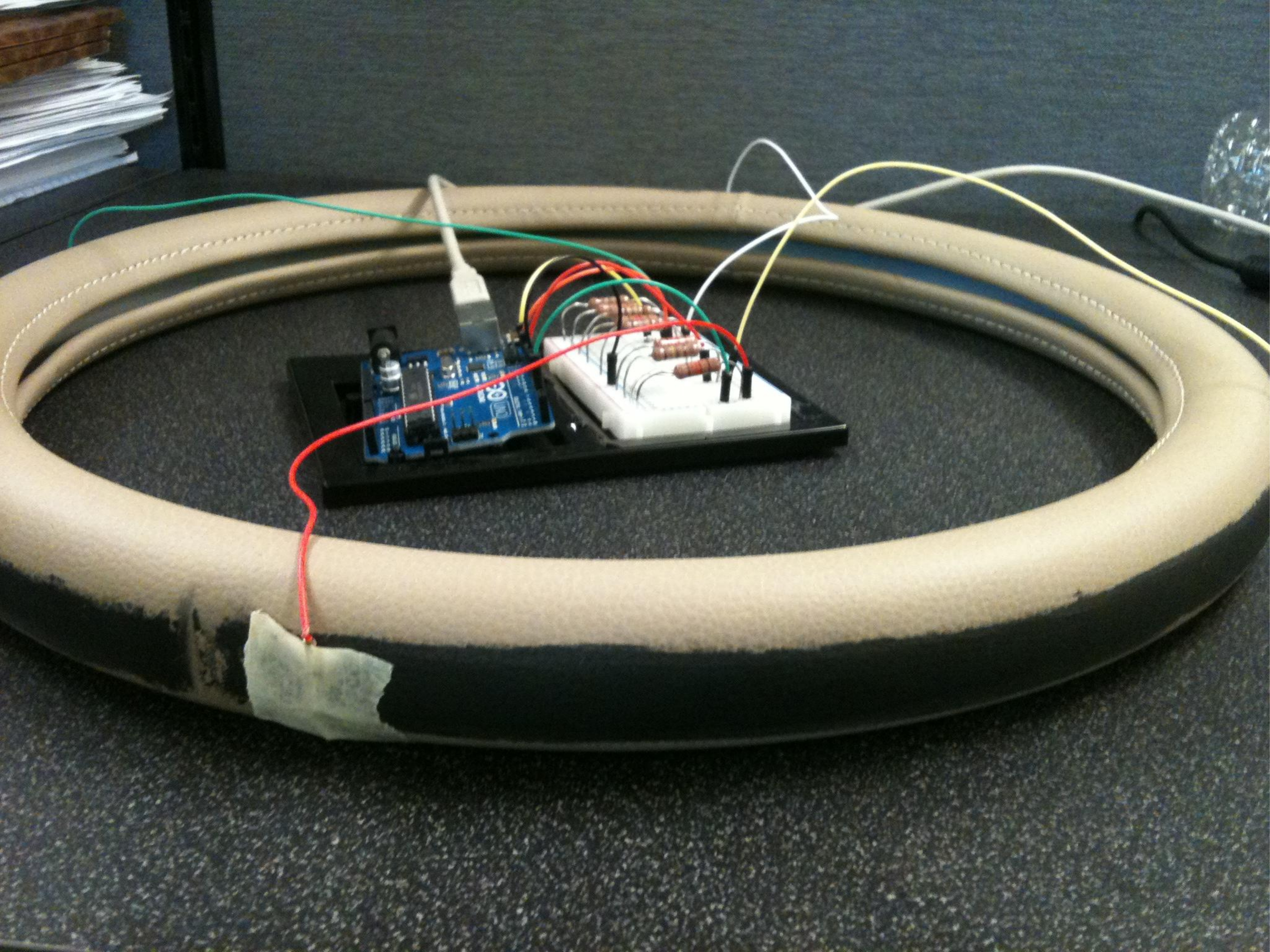}
\caption{\small Touch sensor implemented on a steering wheel cover.  This shows the connections to the Arduino that relays the information over BlueTooth.}
\label{fig:touch}
\end{subfigure}
\caption{\small Capacitive touch sensor for the steering wheel.}
\end{figure}

A touch sensor was built to determine if the driver's hands are on the wheel.  This is done using a capacitive sensor as described in \cite{Bare,Badger}, illustrated in Figure \ref{fig:sensor}, and is implemented using the Capacitive Sensor Library for the Arduino Uno.  The steering wheel is divided into four quadrants, each acting as a sensor.  Conductive paint is used in place of the conductive foil shown in the figure.  Each is connected to an Arduino which continuously sends data over BlueTooth.  The data contains a binary signal indicating whether each sensor is being triggered.  The complete sensor is shown in Figure \ref{fig:touch}.

\begin{figure}[!t]
\centering
\includegraphics[width=0.9\textwidth]{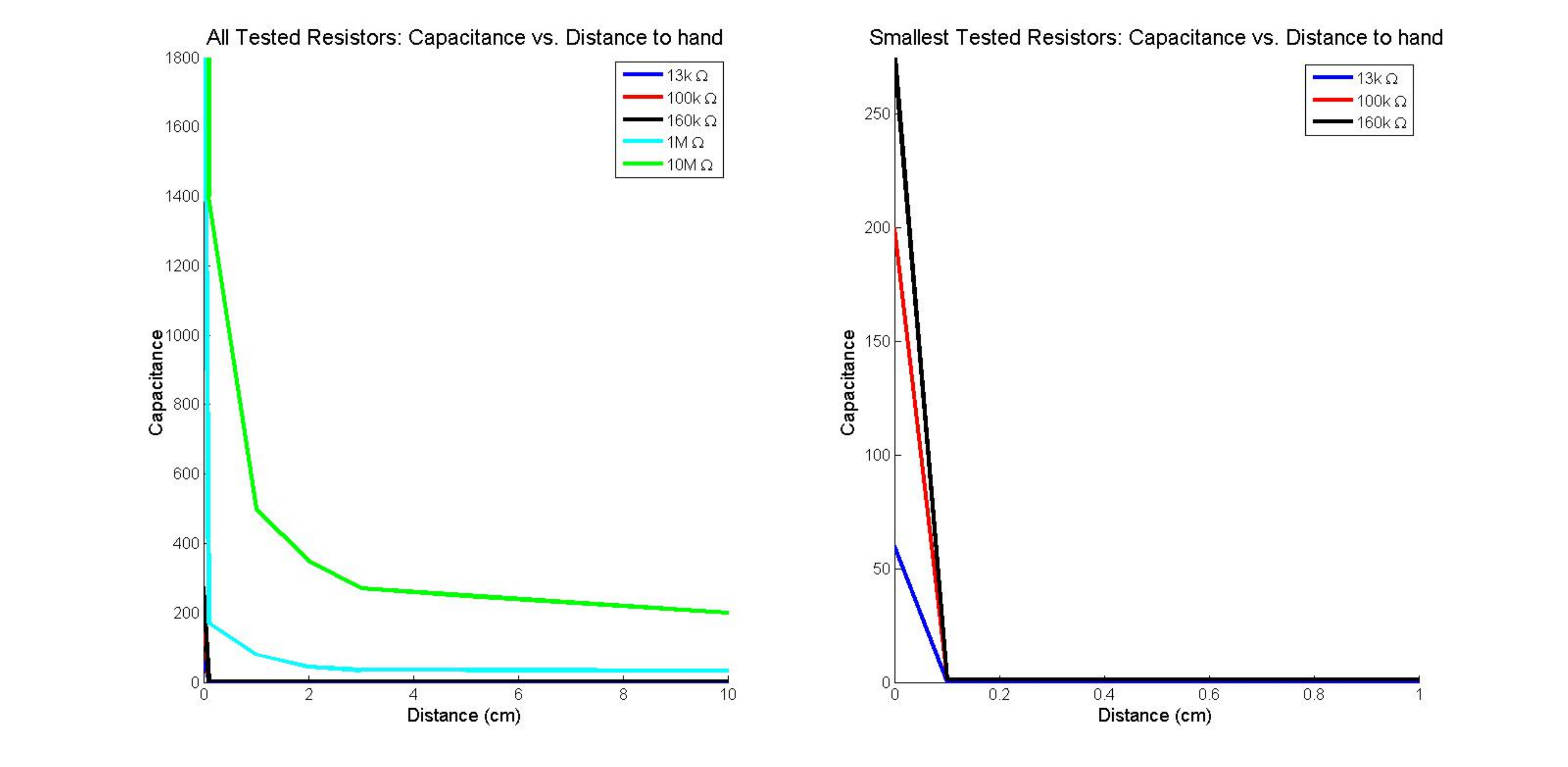}
\caption{\small Calibration plot for the touch sensor showing the capacitance value collected as a function of distance from contact to hand with various resistor values.}
\label{fig:sensor_calibration}
\end{figure}

The sensor was calibrated with various resistor values to examine the sensitivity of the system (Fig. \ref{fig:sensor_calibration}). For calibration, five different resistor values were tested.  The output signal was examined when a hand was placed at six different distances away from the sensor, ranging from zero to ten centimeters.  The $13k \; \Omega$ resistor value was used in the final design, as it resulted in the most boolean response in the calibration process.  This means that it gave no response unless the hand was touching the wheel, while other resistor values gave slight responses if the hand was near the conductive material.

\begin{figure}[!h]
\centering
\includegraphics[width=0.9\textwidth]{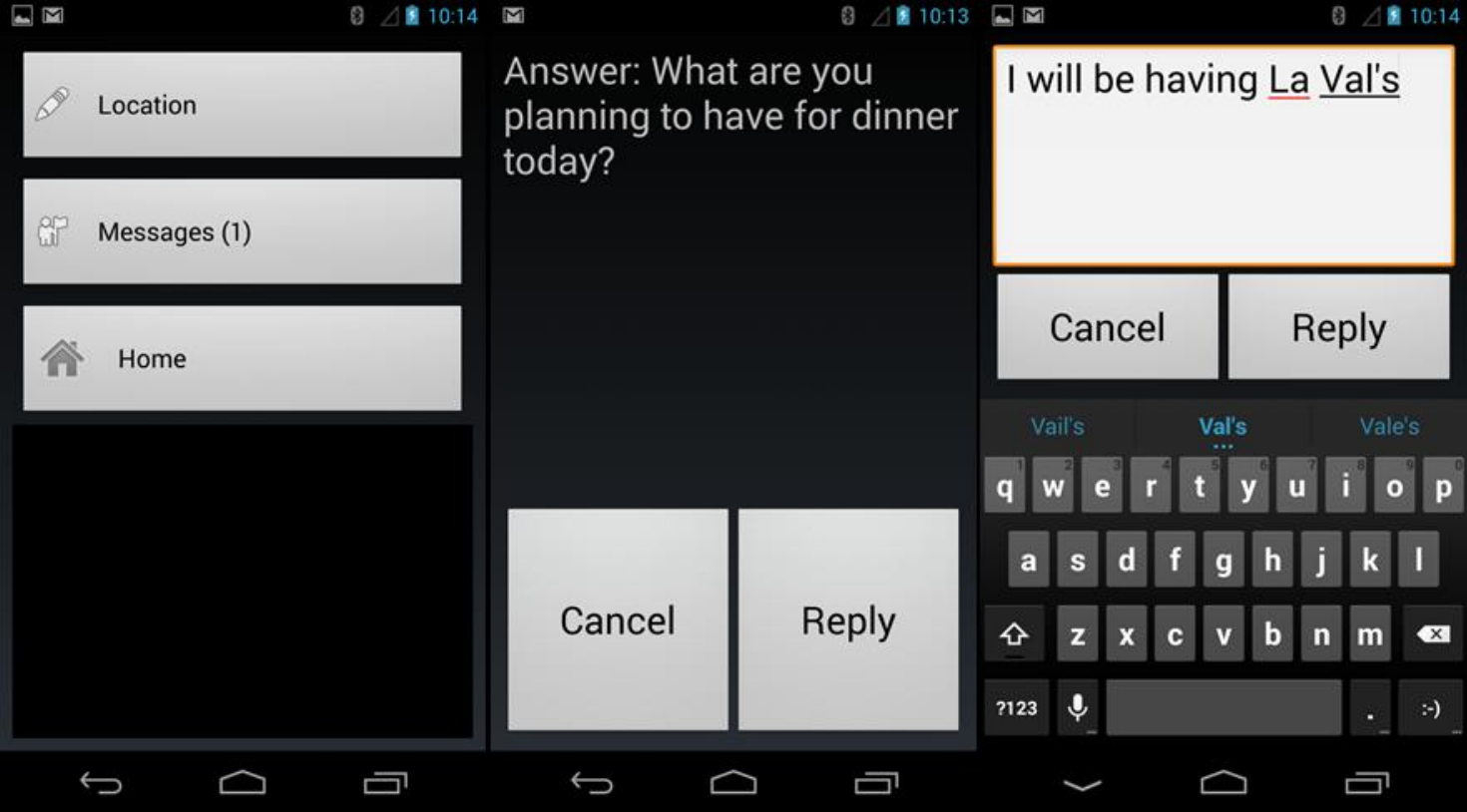}
\caption{\small Android app created to simulate texting environment with randomized messages.}
\label{fig:texting}
\end{figure}

The last additional component added to the experimental design was a phone app to simulate texting, as shown in Figure \ref{fig:texting}.  This application is meant to distract the driver in the form of incoming texts. The application randomly rings in a time frame of 30-60 seconds with a question for the driver to respond to. Upon ringing, this information is sent via BlueTooth to the control server indicating that the user will soon be distracted.  This app can detect when the driver picks up the phone via abnormal changes in the phone's accelerometer data and when the driver starts touching the phone.

\pagebreak

The touch sensor data and the phone app data are sent through the phone to the main computer.  The touch sensor data is sent every 10 ms, while the phone data is event driven based on if the user is picking up the phone, touching the screen, or putting down the phone.

\subsection{Modular Setup}

Once the real-time aspect was completed, the actual design of experiments was considered.  To assist in design and data collection, experiment and data processing modules have been developed.  Simulink modules were created to store all input and output data from the simulator and PreScan.  The components of PreScan that have been set up are briefly explained below.
\begin{description}
\item[Driver Inputs]  The input component is pulled from the simulator to be used as the input to the vehicle model as provided through PreScan.  This includes the braking, throttle, and steering input.

\item[Vehicle Information] The vehicle component provides all vehicle states in addition to GPS coordinates.  These states include: $x,y,$ and $z$ position, rotation on the $x,y,$ and $z$ axes, velocity, heading angle, and yaw rate.  Note that these rotation values are sent to the simulator to recreate the feeling of driving.

\item[Radar Components] The radar component has three different sensors implemented, for the sides and the front of the vehicle.  In PreScan, this is an idealized sensor, so the output for each sensor is the readings for all objects in the experiment world at each time step.  This gives the object ID number (which is determined randomly in PreScan), the angle of detection, the elevation angle of detection, the distance to the object, the velocity of the object, and the heading angle of the object.

\item[Lane Marker Sensor] This component provides all road data at the vehicle and at three predefined distances ahead of the vehicle.  The data collected can provide distance to the nearest lane marker and curb on either side of the vehicle as well as the curvature of the road.  Currently, this is the only data from the sensor being used, although many more measurements can be observed \cite{PreScanMan}. 
\end{description}
These components were selected to mimic the hardware that can readily be implemented on real vehicles, for future testing.  Other components that could be integrated into the experiments can be found in the PreScan manual \cite{PreScanMan}.

\begin{figure}[!b]
\centering
\includegraphics[width=\textwidth]{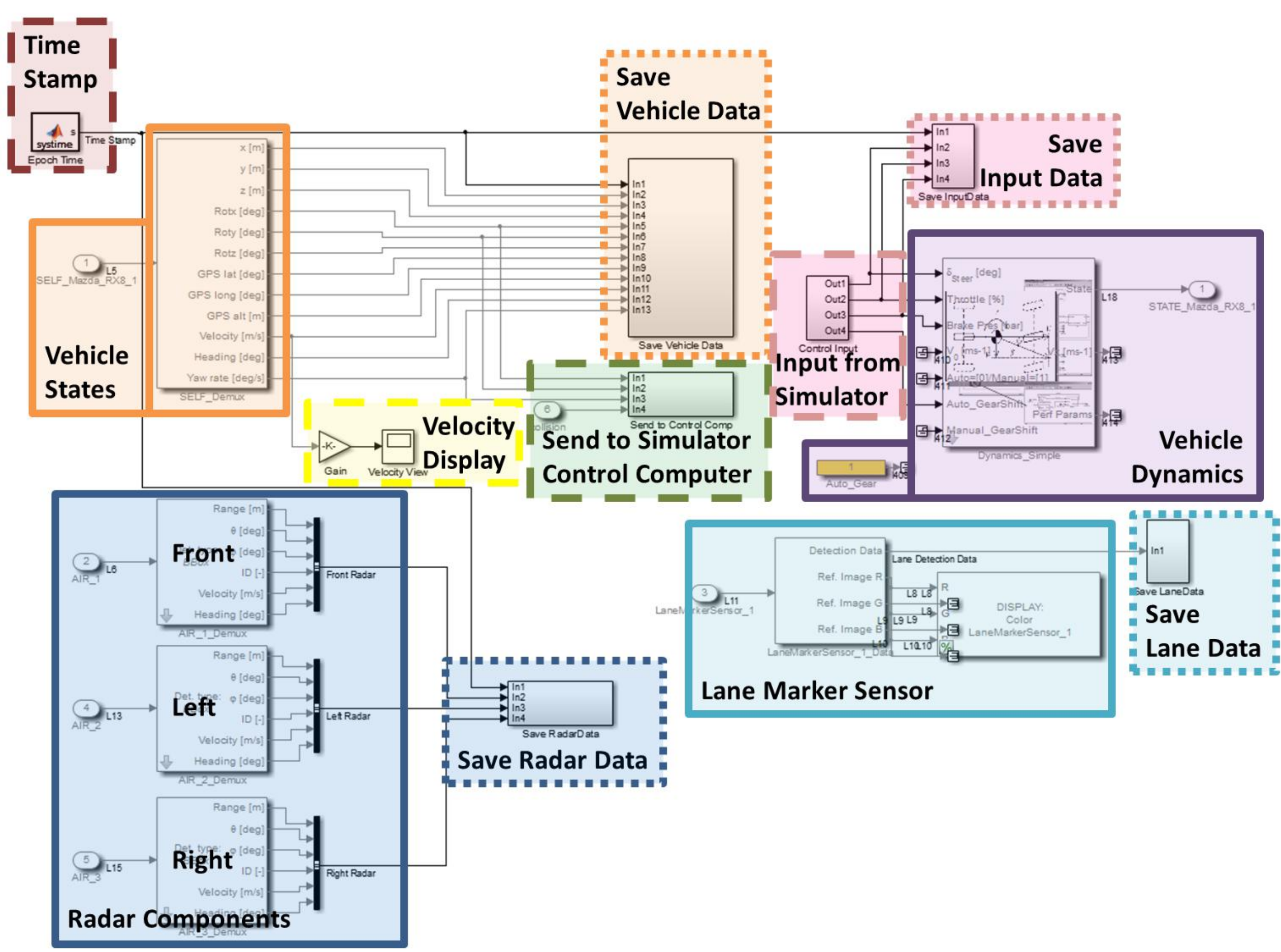}
\caption{\small Schematic of the Simulink setup with labeled components. All components that should be added to the Simulink file (i.e. not provided by PreScan) are shown in dotted lines.  The data collection modules distinguished by the smaller dotted line. Similar colored boxes denote that that they are related to the same component.}
\label{fig:simulink}
\end{figure}

The vehicle, input, radar, and lane marker components each have a data collection module associated with it entitled ``Save $<$component name$>$ Data."  Simulink modules have been made for all the variables that effect the driving experience as well.  This allows for control over the magnitude of the forces the driver will feel while driving.  The vehicle performance can also be modified by changing the gains on the braking and throttle inputs.  Additionally, the simulator is setup to shake if the driver collides with an obstacle in the experiment.  While it is recommended that this vibration is little more than a gentle shake, the magnitude of the shake is also adjustable.
The complete Simulink setup with the components marked is shown in Figure \ref{fig:simulink}.

\pagebreak

\subsubsection{Data Processing}
\begin{figure}[!b]
\centering
\includegraphics[width=\textwidth]{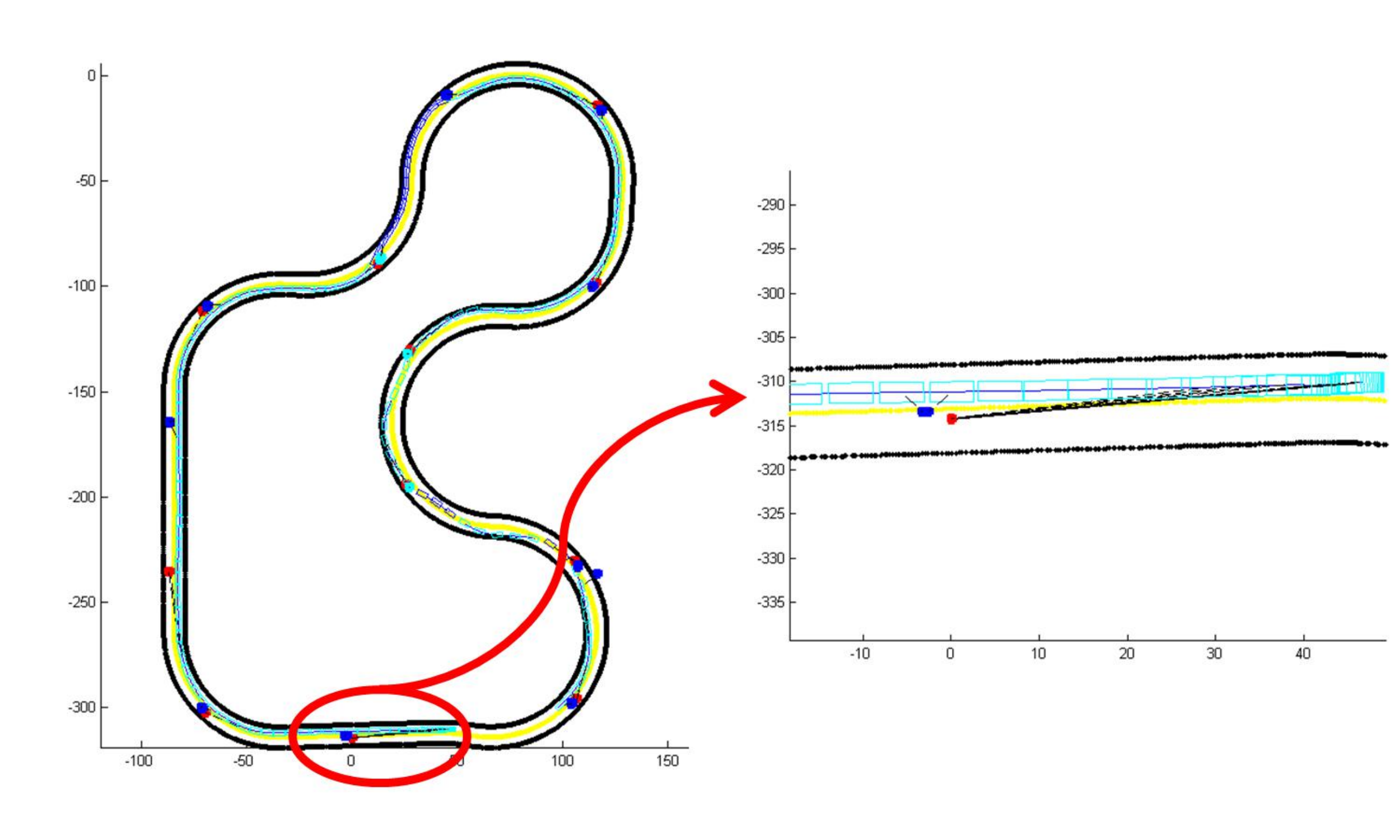}
\caption{\small Matlab plot showing the experimental run.  The road and lane boundaries are shown in black and yellow.  The path the vehicle took is shown in a blue line and the blue or light blue box.  The color of the vehicle box indicates the lane.  Objects are shown in red, blue, or light blue, depending on if the front, left, or right radar detected the object, respectively.  The angle and line of detection is shown from the vehicle.}
\label{fig:plots}
\end{figure}

Matlab scripts are available to easily unpack and do minor processing of the data.  Basic functions can unpack the vehicle, input, radar, and lane marker data.  The lane and vehicle data can be used to determine indicators of which lane the driver is in and offset relative to the center of the road.  The radar data can be interpreted into three components: front object, left object, and right object, giving the data for the closest obstacle.  With this basic processing, the investigator can develop algorithms to assess driver behavior.  The collected data can be plotted to visualize the data collected in the experiment, as shown in Figure \ref{fig:plots}.

\subsection{Safety Measures}
In order to guarantee safety for the subjects, an IRB Protocol was submitted and conditionally approved as of December 6, 2013.  Subjects will only be selected if they are over the age of 18, are not vulnerable to undue influence, and are not pregnant, and will sign a form to acknowledge that they are qualified to perform the driving tasks and are aware of the experimental procedures.  Assuming the subject is rational and will not intentionally attempt to injury himself, the likelihood of physical injury from use of the force dynamics simulator is minimal. For safety, the simulator will not start if the entry gate is not closed or if the seat belt is not on. Also, there are two big red emergency kill-switches: one on the simulator and one remote which will be manned by the experiment proctor.  To minimize risk, the subjects will be instructed on how to stop the simulation at any time before the experiments.  Also, a waist high fence surrounds the simulator while it is active to avoid observers getting hurt.  A curtain is also to be installed around the simulator to avoid unnecessary distractions to the driver.  To avoid potential mental distress, only animals, vehicles, or other inanimate objects will be used as obstacles.  Upon request, the IRB Protocol can be provided for more detailed information.

\section{Discussion and Future Work}
This paper presented the work that has gone into setting up an experimental platform to test Human-in-the-Loop driving experiments.  This has been done using a new force feedback simulator that allows for a realistic driving experience, while allowing complete control over the testing environment.  By utilizing the PreScan software, many variables can be controlled including road configuration, weather, obstacles, vehicle dynamics, and much more.  Combining these two components to form a complete set up has been completed in addition to extra features.  These include driver monitoring devices (including cameras, Kinect, etc.) that can be mounted to the simulator as desired, eye tracking glasses, a touch sensor for the wheel, and a phone app to simulate distracted driving.  To make experiments easy and quick to set up, modules have been created for data collection and for simple data processing that can be used by those interested in running driving experiments.  In addition, safety measures have been implemented in order to guarantee the subject's safety when using the setup.  While this setup is to be used for many experiments, the primary focus will concern human interaction with semi-autonomous or autonomous vehicles.  This can be used to build driver models to build smart active safety systems, as demonstrated in \cite{Shia2013,Vasudevan2012}.

\pagebreak

\section*{Acknowledgements}
Funding for these studies and the simulator was granted by the Department of Defense Office of Naval Research, Award Number N00014-13-1-0341: Embedded Humans: Provably Correct Decision Making for Networks of Humans and Unmanned Systems.


\footnotesize
\begin{thebibliography}{9}

\bibitem{Bare}
Bare Conductive. ``Making a Capacitive Proximity Sensor with Bare Paint." \url{http://www.bareconductive.com/capacitance-sensor/}
\bibitem{Badger}
Paul Badger. ``Capacitive Sensing Library," Arduino Playground. \url{http://playground.arduino.cc//Main/CapacitiveSensor?from=Main.CapSense/}
\bibitem{ForceDynamics}
Force Dynamics - 401 Simulator. \url{http://www.force-dynamics.com/}
\bibitem{Greenberg1994}
Greenberg, J. and Park, T., ``The Ford Driving Simulator," SAE Technical Paper 940176, 1994.
\bibitem{DriggsCampbell2013}
K. Driggs-Campbell, V. Shia, R. Vasudevan, F. Borrelli, and R. Bajcsy. ``Probabilistic driver models for semiautonomous vehicles," in Digital Signal Processing for In-Vehicle Systems. Seoul, South Korea. October 2013. 
\bibitem{OSUsim}
Driving Simulation Laboratory at The Ohio State University. \url{http://drivesim.osu.edu/facility/}
\bibitem{ClemsonSim}
Driving Simulator Lab, Department of Psychology at Clemson. \url{http://www.clemson.edu/psych/research/labs/driving-simulator-lab/}
\bibitem{Trivedi}
Laboratory for Intelligent and Safe Automobiles, Intelligent Vehilce novel Experimental Test Beds. \url{http://cvrr.ucsd.edu/LISA/TestbedThrust.htm/}
\bibitem{UCSDsim}
Language \& Cognition Lab at the University of California, San Diego. Driven to distraction. \url{http://www.cogsci.ucsd.edu/spotlight/9/}
\bibitem{PreScan}
PreScan, TASS International. A Simulation and Verification Environment for Intelligent Vehicle Systems. \url{http://www.tassinternational.com/}
\bibitem{DoD}
Secretary of Defense for Acquisition Technology. DoD Modeling and Simulation Glossary.  \url{http://www.dtic.mil/whs/directives/corres/pdf/500059m.pdf/}
\bibitem{Shia2013}
V. Shia, Y. Gao, R. Vasudevan, K. Driggs-Campbell, T. Lin, F. Borrelli and R. Bajcsy, ``Driver Modeling for Semi-Autonomous Vehicular Control,” IEEE Transactions on Intelligent Transportation Systems, Under Review, 2013.
\bibitem{SMI}
SMI Eye Tracking Glasses. \url{http://www.eyetracking-glasses.com/}
\bibitem{PreScanMan}
TASS International. PreScan Manual, Chapter 5: Experiment Components. Printed May 9, 2013.
\bibitem{ToyotaSim}
Toyota Research. Pursuit for Vehicle Safety: Driving Simulator (Safety Technology Innovation from a Driver's point of view). \url{http://www.toyota-global.com/innovation/safety_technology/safety_measurements/driving_simulator.html}
\bibitem{Vasudevan2012}
R. Vasudevan, V. Shia, Y. Gao, R. Cervera-Navarro, R. Bajcsy, and F. Borrelli. ``Safe Semi-Autonomous Control with Enhanced Driver Modeling." In American Control Conference, 2012.
\bibitem{DMV}
Vermont Department of Motor Vehicles: Agency of Transportation.  ``Follow the 4-Second Rule for Safety Spacing." \url{http://dmv.vermont.gov/sites/dmv/files/pdf/DMV-Enforcement-SM-4_Second_Rule.pdf/}
\bibitem{StanfordSim}
Volkswagon Automotive Innovation Lab at Stanford University. \url{http://www.standford.edu/group/vail/}

\end{thebibliography}
\end{document}